\documentstyle[12pt]{article}
\input psfig.sty
\def\singlespace{\baselineskip=13.5pt\lineskip=0pt
\lineskiplimit=-5pt}
\def\title#1{\relax\vspace*{2cm}{\large{\bf #1}}\par\vspace*{13.5pt}}
\def\author#1{{#1}\par\vspace*{13.5pt}}
\def\affil#1{{\it #1}\par}
\def\abstract{\vspace*{27pt}ABSTRACT\par\relax}
\def\section#1{\par{#1}\par}
\def\subsection#1{\par\underline{#1}\par}
\def\subsubsection#1{\par\underline{#1.}\ \ }
\def\acknow{\par ACKNOWLEDGMENTS\par}
\newenvironment{references}{\section{REFERENCES}\vspace*{.5cm}%
\parindent=0pt\frenchspacing%
\parskip=1pt plus 1pt minus 1pt%
\interlinepenalty=1000\tolerance=400%
\pretolerance=10000\hyphenpenalty=10000%
\everypar={\hangindent=1.6pc}
}{}

\def\ltwid{\mathrel{\raise.3ex\hbox{$<$\kern-.75em\lower1ex\hbox{$\sim$}}}}
\def\gtwid{\mathrel{\raise.3ex\hbox{$>$\kern-.75em\lower1ex\hbox{$\sim$}}}}

\begin{document}
\singlespace
\title{MILLIMETER-WAVE SPECTROSCOPY AND MAPPING OF QUASAR HOSTS, AND THE STATUS
OF ULIRGS AS QUASAR 2S}
\author{Robert Antonucci}
\affil{University of California, Santa Barbara\\}

\abstract
It is becoming possible to detect high redshift quasars in various
molecular lines, and to show by mapping lensed objects that the strong
dust and molecular emission arises in warm dense $\sim$ 100 pc-scale ``tori."
The properties of ULIRGs, at least those with AGN-like narrow line 
regions, are very similar, as expected in the hidden quasar hypothesis.
Several of the latter are in fact confirmed as ``Quasar 2s" by 
spectropolarimetry.

\section{INTRODUCTION: DETECTIONS OF MOLECULES IN QUASAR HOSTS}

Millimeter-wave observatories are just now reaching the sensitivities and
spatial resolutions required for studying nuclear molecular tori and host
galaxies associated with quasars.  In the late 1980s and early 1990s, the mm
and submm dust continuum was detected in some low-redshift ($\sim 0.1$) quasars
(e.g. Chini et al. 1989, Barvainis et al. 1992).
Subsequent spectroscopy showed strong CO emission, as expected for luminous
dusty sources (e.g. Sanders et al. 1988; Barvainis et al. 1989b; Alloin et al.
1992).

In general the detections and limits similar to those for ultraluminous
infrared galaxies at the same far-IR luminosity \footnote{Kennicutt (1990) has extended the L(CO) - L (FIR) correlation
for galaxies by adding points for a burning cigar, a Jeep Cherokee, the 1988
Yellowstone National Park forest fire, and Venus.  I think the lesson is you
get correlations in luminosity - luminosity plots (even without any sample    
incompleteness) just because big things have more of everything.  There is
probably a good correlations between the number of bookstores in a city and the
number of bars, even with a complete sample, but no direct connection is
necessarily implied.  However, there is much other evidence that the IR continuum in 
quasars
is dust emission (e.g. Barvainis 1992), and that it is intimately related to
the molecular emission (e.g. Alloin et al. 1992.)}(Alloin et al. 1992, 
Figure 3). 
This supports the notion that 
ultraluminous IRAS galaxies harbor hidden
quasars.  Spectropolarimetric observations have shown this to be true in many
cases, and furthermore that the quasars are not ``buried" but are visible from other
directions.

\section {DETECTIONS AT HIGH REDSHIFT}

Following breakthrough detections of the ultraluminous IRAS galaxy F10214+4724  
(Brown and vanden Bout 1992, Solomon et al. 1992b), our group studied the
Cloverleaf quad lens BAL quasar in various lines (Barvainis et al. 1994, 1997),
including deep high resolution mapping in the CO 7-6 line (Alloin et al. 1997,
Kneib et al. 1998;
See Yun et al. 1997 for an earlier OVRO image).

The last two years have brought detections of BR 1202-0725 at Z=4.7 (Ohta et
al. 1996, Omont et al. 1996), 53W002 at Z=2.3 (Scoville et al. 1997), MG 0414+0534 at Z=2.6,
(Barvainis et al. 1998), BRI 1335-0415 at Z=4.4 (Guilloteau et al. 1998), and
APM 08279+5255 at Z=3.9 (Downes et al. 1998).

Two factors have enabled the detection and study of high-redshift dust and
molecular emission, aside from rapid increases in instrumental sensitivity and
resolution.  The K-correction term is so favorable for this situation that the
continuum and line fluxes are both flat or increasing with redshift!  And in
most of the detected objects the fluxes are greatly boosted by gravitational   
lensing.  Prospective improvements in instrumental sensitivity can do the
amplification job that lensing does so that soon the early universe of (relatively) ordinary objects
will be revealed (e.g. Barvainis 1996); it will take a little longer to do
without the lensing magnification of the angular sizes, required for resolving
much of the emission.
    
\section {SUMMARY OF MILLIMETER OBSERVATIONS OF THE CLOVERLEAF QUAD LENS
QUASAR} 

All of the high-Z CO detections warrant follow-up observations of other
millimeter lines in order to evaluate the molecular and atomic masses as well
as the physical conditions. 
Mapping with reasonable velocity resolution also provides a wealth of
information on morphology and dynamics.  These follow-up observations are in
advanced stages only for F10214+4724 at Z=2.3 and for the Cloverleaf quad at
Z=2.6.   

When F10214+4724 was first discovered, because of its incredible reported
luminosity and the simultaneity with a certain geopolitical event, I started
thinking of it as the Mother of All IRAS Galaxies.  The original reported
CO J=2-1 luminosity,  in Brown and vanden Bout 1991, is L$_{\rm CO}$ = 
1.9$\times 10^{9}h^{-2} L_{\odot}$.  This is stated to be 15,000 times the
Milky Way value.  The corresponding H$_2$ mass would then be $1.8 \times
10^{13} h^{-2}L_{\odot}$ for q$_0$=1/2, if the conversion factor is similar to that
of Galactic molecular clouds.  However, since then the CO flux (Radford et al. 1996),
the luminosity distance (Solomon et al. 1992a), and probably the CO-to-H$_2$
conversion factor (see e.g. Solomon et al. 1997, Barvainis et al. 1997) have
been revised downwards substantially; also the lensing amplification was
recognized and corrected (e.g. Radford et al. 1995, Green and Rowan-Robinson
1996).  Thus the point for F10214 or the L(CO) vs L(FIR) plot of Kennicutt 1990
has now moved to just above the point representing the Jeep Cherokee.
                                                                                
The bulk of the CO emission in F10214 comes from a $\sim 1$" source (e.g.
Downes et al. 1995, Scoville et al. 1995) and the same is true for the
Cloverleaf (Yun et al. 1997, Alloin et al. 1997, Kneib et al. 1998).

The best mapping data on the Cloverleaf CO is that of Kneib et al.; it has
better surface brightness sensitivity, especially to $\gtwid 1$" extended emission, than
Alloin et al., and substantially better sensitivity and resolution than Yun et
al.  The maps detect the four images very well, and resolve each one
individually. \footnote{This type of data is helpful for constraining
macrolensing parameters.  Unlike in the optical, the mm images are unaffected by
variability, microlensing, and foreground extinction.}
As interpreted by Kneib et al. with a detailed and fairly well-constrained lens
model, the source must be intrinsically quite small - only $\sim 100$ pc in
radius for $h=0.5, \  q_0 = 1/2$. (This size constraint and the observed flux
requires that $T_{B} \gtwid 40$K; the $T_B$       
constraint deduced from modeling below requires that the radius is at least 40pc.)
It is possibly rotating, although an unrelaxed merger is also a plausible
interpretation.  The interior dynamical mass is of order $10^{9}h^{-1}M_\odot$.  
This compares with a deduced molecular and atomic mass of a few times
$10^{10}m^{-1}h^{-2}M_{\odot}$, where m is the lens amplification (Barvainis et
al. 1997), and a minimum black hole mass of $\sim 1 \times 10^9 m^{-1}
h^{-2}M_\odot $ by the Eddington limit argument.  New observations of another
dusty lensed BAL at even higher redshift, APM 08279+5255 at Z=3.9, paint a very
similar picture for that object (Downes et al. 1998, Ledoux et al. 1998).

With this size scale and gas mass, we are clearly talking about nuclear gas
rather than the distributed gas of a protogalaxy.  This gas may be analogous
to that seen in nearby luminous AGN such as NGC 1068 (Planesas et al. 1991,
Jackson et al. 1993, Tacconi et al. 1994).  In both cases it probably plays
the role of the (outer part of the) dusty torus invoked in Unified Models for
AGN (Antonucci 1993).  The spectral energy distributions of the Cloverleaf and  
F10214 are remarkably similar in the far-IR, with the differences very well
explained as resulting from absorption and scattering (Barvainis et al. 1995).
The scattered UV light in F10214 reveals a hidden quasar (Goodrich et al.
1996), so it seems certain that that object appears as a quasar from some other
directions in space.  Again this is generally consistent with a simple and
far-reaching Unified Model, in which many ultraluminous IRAS galaxies are simply
``Quasar 2s," the high luminosity extension of the Seyfert 2s.  The optical
light from F10214 is mainly from the reflecting mirror or periscope familiar from
studies of local Seyfert 2s, other luminous IRAS galaxies, and distant narrow
line radio galaxies.

Our group has also made a rather detailed multi-line observational and
theoretical study of the integrated emission from the Cloverleaf (Barvainis et
al. 1997).  Four lines from the CO ladder have been observed well: J = 3-2, 4-3,
5-4 and 7-6.  The line brightness temperatures seem to rise with J and then fall
off, suggesting warm dense gas of only moderate optical depth.  This would    
imply high effective emissivity and thus a fairly low molecular mass for the
observed CO luminosity.  Since systematic observational errors are still
possible, the relatively low optical depths shouldn't be considered proven.
($^{13}$CO might be detectable and provide a more reliable molecular mass.) 
The strong CI line implies a large atomic mass, larger than that of H$_2$
(assuming the moderate optical depths for CO).  An HCN detection needs
confirmation, but is at the level expected based on ultraluminous IRAS
galaxies; of course it requires a considerable mass of gas which is quite
warm and dense.  The physical conditions and masses inferred from this
multi-line study are in accord with the 100-pc scale nuclear ``torus" suggested
by the direct imaging.

\section {RELATIONSHIP OF ULTRALUMINOUS IRAS GALAXIES TO QUASARS}

In the 1970s it was commonly held that there must be few if any ``Quasars 2s,"
that is, objects with just narrow emission lines, but as luminous as quasars.  
But the Seyfert 2 optical continua are generally dominated by the light from
old stellar populations, so they wouldn't scale up with luminosity of the
central engine.  It's true that Quasar 2s would have luminous narrow lines, but
the narrow line equivalent widths go down as luminosity increases in Type 1
AGN, so the NLR luminosities of the Quasar 2s might not be huge. The point is
that Quasar 2's would not be so easy to find.

In the Unified Model, claimed to be correct ``to zeroth order" (Antonucci 1993), all
AGN are surrounded by obscuring tori which probably have dust opacity
dominating in the optical region of the spectrum.   These hide the nuclear continuum
and broad line region in the 2's.  The waste heat from the tori appears in the
IR, and while it may not scale exactly with luminosity if the torus opening angle is
changing, it should still be conspicuous  in Quasar 2s.  This is strongly supported
by the rather well established dust emission in quasars: it accounts for 
$\sim 30$ \% of the luminosity in UV-selected objects (e.g. Sanders et al. 1989).  This
further implies that if the tori are opaque and if they  produce most of    
the FIR (as is very likely in the Cloverleaf, for example), then the torus
covering factor is typically $\sim 30$\%; thus there should be twice as many Quasar 1s as
2s at a given value of an isotropic parameter such as far-IR (but {\it not}
bolometric) luminosity.  A Quasar 2 should also show Seyfert-like narrow lines
ratios, with strong emission
from low-ionization species such as [OI] and [NII], and also from high ionization
species such as [OIII].

These expectations are borne out for a large fraction of ultraluminous IRAS
galaxies. At the highest luminosities at least half have Seyfert 2-like narrow
line ratios (e.g. Kim et al. 1998; see also Surace and Sanders 1998 for the
finding that all twelve ULIRGs imaged at $2.1 \mu$ show strong point sources
which they attribute to hidden AGN).  
In many cases, the polarized flux spectrum has been measured and it
shows the Type 1 nucleus.  The space densities of ultraluminous IRAS galaxies
and quasars are approximately consistent with the above expectations         
(Gopal-Krishna and Biermann 1998).  The same is true on the radio loud side,
with a large fraction of the most luminous 3C radio galaxies already shown to
be quasars in polarized flux, with polarization position angle indicative of
polar scattering as in the Seyfert 2s, and consistent with the Unified Model. 
Again at least for the most luminous and distant objects, the space densities of Type 1 and
Type 2 radio loud AGN are consistent with expectations (Barthel 1989).

What I'd like to do with the remainder of this section is to discuss some
issues which have arisen in the literature recently, attempting to be as
disagreeable as possible.

Let's consider whether ISO and ASCA have anything to say whether ultraluminous
infrared galaxies contain optically obscured AGN, and whether AGN or starbursts
contribute more than 50\% of their power.  In general I think it is an
unjustified assumption that you can see through to the nucleus in the mid-IR. 
Soon after the polarized light trick revealed hidden Type 1 nuclei in Seyfert    
2s and narrow line radio galaxies (results published in 1982-1985), it was
recognized that the nuclear x-rays in the Type 2s must be blocked by the tori
(Antonucci 1984), and that the tori are thus Compton
thick in many cases (Krolik and Begelman 1986).  The lack of broad Br$\alpha$ in
total flux in NGC1068 provides a similar constraint (A$_\nu \gtwid 1000$). 
Molecular mapping of NGC1068 (references given above), and for example Arp220
(Scoville,
this meeting), confirm it.

More recently x-ray spectra (many papers by Awaki, Koyama and others) showed
columns of $\sim 10^{23}$cm$^{-2}$ to $> 10^{24}$cm$^{-2}$, including some
Compton-thick cases.  Maiolino et al. (1998) argue persuasively that these earlier
studies selected preferentially the {\it lowest} column densities by selecting
objects with known polarized broad lines.  The latter are known to have relatively
highly inclined tori (warm IR colors) in the Unified Model (e.g. Heisler et al.
1997 and several important references therein).  In any case the {\it other } bright 
Seyfert 2's have much larger columns with many Compton thick examples found
(Maiolino et al. 1998). 
Therefore, going to the mid-IR does not penetrate such tori in
general. \footnote{It
follows that the $12\mu$-selected sample studied by Malkan and collaborators,
designed to be isotropic and still seemingly claimed to be so, is in fact
highly anisotropic.  See e.g. Rush et al. (1996).}
There
may be cases in which a high sight line or a ``nonstandard" dust distribution
allows such penetration, but it certainly shouldn't be generally assumed.  In fact if the
tori contain star forming regions,  mid-IR spectra should show HII-region line
ratios.  

One can make quantitative (though model-dependent) arguments that the power in
certain mid-IR emission lines implies a certain total starburst contribution to
the luminosity.  But in order to prove the starburst contributes {\it most } of         
the luminosity (which  means $> 50\%$) you'd need to show that the observed      
feature correlates with starburst
bolometric luminosity with a very small
dispersion.  Similarly a Seyfert-like x-ray source proves that the AGN
{\it dominates} the bolometric luminosity only if the two can be shown to correlate
with a sufficiently small dispersion.

Of course ISO spectra can and do sometimes reveal Seyfert-like narrow line
regions in some Compton thick AGN for the same reason optical spectra do: the
NLR extends above the torus.  In fact they seem to be the same objects - so
it's not clear what ISO has added to this particular question.

It is sometimes said that the mid-IR colors of many ULIRGs are too cool 
for hidden AGN, but this assumes a
small outer radius for the tori.  It is argued by Genzel et al. (1998) that
larger tori are implausible, or tightly constrained by molecular line
observations in some cases.  For Seyfert 2s, Maiolino et al. (1995), Section   
4.3, similarly find that the Type 2s are cooler than expected, though in that
case they have first removed the warm ones preferentially by excluding those
with easily  detected Type 1s in polarized flux.

G. Rieke stated at the Tucson meeting on the Galactic Center (Sept. 1998;
proceedings will emerge eventually) that ULIRGs have $L_{BOL}/L_{Eddington}$ up
to $\sim
0.1$ (assuming black  holes with 0.5\% of the bulge mass), versus up to $\sim 1.0$
for quasars.  This was interpreted as showing that ULIRGs are not hidden
quasars.  It does rule out ULIRGs as {\it buried} quasars, where by that I
mean quasars with unity dust covering factor.  But I believe it is just what
we would expect for Quasar 2s.  First in that case $L_{BOL}$ is anisotropic
by a factor of 3 or so just because the big blue bump is obscured in the 2s
(based on the PG-quasar SEDs and the relative space densities, both cited
above). 
And second, $L_{BOL}$ is anisotropic because the IR itself      
is also anisotropic by a
factor of $\sim 3$ according to the models  cited earlier for these extremely
opaque tori.   (I think the mere fact that Seyfert  2's and ULIRGs which would
be seen as type 1s from other directions (the spectropolarimetry argument) have
steeper IR spectra than other type 1's proves the IR is highly anisotropic.)

On another topic, I wonder if there is any evidence for an evolutionary merger 
sequence between
quasars and ULIRGs, as proposed, for example, by Sanders et al. 1989.
In one variant of the idea, using data on ULIRGs only, those dominated by
starbursts would correspond to earlier merger stages than those dominated by
AGN.    Suppose, optimistically, that the few-
keV ASCA spectra and the ISO mid-IR spectra do diagnose correctly by the line
emission which is the primary energy source. (This is fairly similar to
classifying by the optical spectra in any case, as explained above.) The
predicted effect is not seen so far, using ASCA                  
(T. Nakagawa, this meeting), or ISO (Genzel et al. 1998) classifications.

A related hypothesis can accommodate the possibility that all ULIRGs (with Seyfert-like narrow line
spectra) would look like quasars if oriented differently.  It certainly seems
reasonable that the ULIRGs or Seyfert 2s would tend to  have larger covering factors
(smaller torus opening angles) than quasars.  And the dust configurations, with
their large scale heights, may be very much non-equilibrium configurations, and
in fact they are only thought to be geometrical tori ``to zeroth order".  Has anyone
tried to evaluate statistically the relative merger stage of ULIRGs and
quasars?

As a final comment, it is often stated that the light from the nuclei in Seyfert
2s and ULIRGs may be hidden by a warped thin disk or some other configuration
rather than a thick disk.  We do know that in some cases the nucleus is hidden
by dust in the host plane, and {\it not} by a nuclear torus.  This crucial
point was made by Keel 1980, Lawrence and Elvis 1982, de Zotti and       
Gaskell
1985 and many others.  Implications for tests of Unified Models are discussed
in Antonucci 1993, See 2.2, and Antonucci 1998.

But in most cases a thick torus, tipped with respect to the host,
is indicated by a {\it high} optical
polarization oriented perpendicular to a tipped radio jet.  It is a mistake to
assume that the low reported {\it broadband}
polarizations for some 2s are indicative of the polarization of the scattered 
photons. Even after removal of the light from the old stellar populations,  we
know that there is a diluting continuum source, called FC2 by J. Miller and
collaborators (e.g. Tran 1995). It was made clear in all the polarization papers
that 1) the broad line polarizations are very high - usually just a lower
limit is available because the broad line can't be seen clearly in total flux;
and 2) the broad line equivalent widths in the polarized flux spectra are
normal. Thus in every case a normal Type 1 nucleus is revealed in polarized  
flux, and the scattering polarization is very high. One misapplication of the data, claimed to argue against the thick disk
for optical obscuration, is in Malkan et al.  1998; see Antonucci 1998 for
other demurs regarding that paper.

Of course the only sense in which the obscurers are claimed to be 
geometrically thick toris
is that the optical photons can only escape in the polar directions.

\acknow

Most of this work was supported by NSF AST 9617160.  I thank Rich Barvainis for
help preparing this talk, and for comments on the paper.

\end{document}